\documentclass[aps,pre,preprint,groupedaddress]{revtex4}
\usepackage{graphics,epsfig}
\usepackage{amsmath,amssymb}

\begin{document}

\title{Plateaux formation, abrupt transitions, and fractional states in a
competitive population with limited resources}
\author{H. Y. Chan,$^1$ T. S. Lo,$^{1}$, P. M. Hui,$^1$ and
N. F. Johnson$^2$}
\affiliation{$^{1}$Department of Physics, The Chinese University of
Hong Kong\\
Shatin, New Territories, Hong Kong\\
$^{2}$Department of Physics, University of Oxford, Oxford OX1 3PU,
United Kingdom}

\begin{abstract}
We study, both numerically and analytically, a
Binary-Agent-Resource (B-A-R) model consisting of $N$ agents who
compete for a limited resource $1/2 \leq L/N \leq 1$, where $L$ is
the maximum available resource per turn for all $N$ agents.  As
$L$ increases, the system exhibits well-defined plateaux regions
in the success rate which are separated from each other by abrupt
transitions. Both the maximum and the mean success rates over each
plateau are `quantized' -- for example, the maximum success rate
forms a well-defined sequence of simple fractions as $L$
increases. We present an analytic theory which explains these
surprising phenomena both qualitatively and quantitatively. The
underlying cause of this complex behavior is an interesting
self-organized phenomenon in which the system, in response to the
global resource level, effectively avoids particular patterns of
historical outcomes.


\end{abstract}

\maketitle \thispagestyle{empty}

\section{Introduction}
\label{sec:PhaseTrans_Intro}

Complex systems have attracted much attention among physicists,
applied mathematicians, engineers, and social scientists in recent
years. In particular, agent-based models have become an important
part of research on Complex Adaptive Systems
\cite{recentactivities}. For example, self-organized phenomena in
an evolving population consisting of agents competing for a
limited resource, have potential applications in areas such as
engineering, economics, biology, and social sciences
\cite{recentactivities,ourbook}. The bar-attendance problem
proposed by Arthur \cite{arthur1,johnson1} constitutes an everyday
example of such a system in which a population of agents decide
whether to go to a popular bar having limited seating capacity.
The agents are informed of the attendance in past weeks, and hence
share common information, make decisions based on past experience,
interact through their actions, and in turn generate this common
information collectively.  These ingredients are key
characteristics of complex systems \cite{ourbook}.  An important
step in the recent explosion of research in agent-based models
within the physics community, has been the introduction of binary
Ising-like versions of models of competing populations -- examples
include the Minority Game (MG) \cite{challet1,challet0} and the
Binary-Agent-Resource (B-A-R) game
\cite{johnson2,johnsonDPG,zheng2}.

For modest resource levels in which there are more losers than
winners, the Minority Game \cite{challet1,challet7} represents a
simple, yet highly non-trivial, model that captures many of the
essential features of such a competing population.  It has been
the subject of many theoretical studies
\cite{johnson1,challet3,challet7,challet4,johnson2,johnson3,hart1,challet5,challet6,coolen1,coolen2,challet8,galla}.
The MG considers an odd number $N$ of agents. At each timestep,
the agents independently decide between two options `$0$' and
`$1$'. The winners are those who choose the minority option.  The
agents learn from past experience by evaluating the performance of
their strategies, where each strategy maps the available global
information, i.e. the record of the most recent $m$ winning
options, to an action. One important quantity in the MG is the
standard deviation $\sigma$ of the number of agents making a
particular choice. This quantity reflects the performance of the
population as a whole in that a small $\sigma$ implies on average
more winners per turn, and hence a higher success rate per turn
among the agents. In the MG, $\sigma$ exhibits a non-monotonic
dependence on the memory size $m$ of the agents
\cite{savit,challet3,challet4}. When $m$ is small, there is a
significant overlap between the agents' strategies. This crowd
effect \cite{johnson2,johnson3,hart1} leads to a large $\sigma$,
implying the number of losers is high. This is the crowded, or
informationally efficient, phase of MG. In the informationally
inefficient phase where $m$ is large, $\sigma$ is moderately small
and the agents perform better than if they were to decide their
actions randomly.  In this regime, information is left in the
resulting bit-string patterns for a single realization of the
system.  In the inefficient phase, the MG can be mapped on to
disordered spin systems and hence the machinery in statistical
physics of disordered systems, most noticeably the replica trick,
can be applied \cite{challet5,challet6,coolen1,coolen2,challet8}.
However, the replica trick becomes ineffective in the efficient
phase. The Crowd-Anticrowd theory
\cite{johnson1,johnson2,johnson3,hart1,ourbook} gives a physically
transparent, quantitative theory of the observed features of the
MG in both the efficient and inefficient regimes. The
Crowd-Anticrowd theory is based on the fact that it is the
difference in the numbers of agents playing a given strategy $R$
and the corresponding anti-correlated strategy $\overline{R}$,
that dictates the size of the fluctuations and hence performance
of the population as a whole.

In agent-based models, the non-triviality of the results comes
from the actions taken by the agents which are directly related to
the decision mechanism. The decision mechanism depends sensitively
on how each strategy performs at the moment of decision.  In the
efficient phase of the MG, no strategy outperforms the others and
therefore the relative performance of the strategies oscillates as
the game proceeds. This anti-persistent nature of the strategy
performance
\cite{savit,challet3,challet4,larry1,zheng1,Jefferies1,Lo2004} is
crucial in arriving at a quantitative understanding of the MG's
dynamics. As the systems evolves, it goes from one $m$-bit outcome
or history bit-string to another.  Mathematically, the evolution
can be viewed in terms of transitions in the global information
(i.e. history) space.  The $2^{m}$ possible history bit-strings
for a given value of $m$ constitute the nodes in this history
space. As the system evolves, it makes transitions from one node
to another. Jefferies {\em et al.} showed that in the efficient
(i.e. low $m$) phase of the MG, an effective restoring force
dominates the strategy-score dynamics yielding a Eulerian Trail
quasi-attractor in history space \cite{Jefferies1}. As $m$
increases, a competing bias term -- associated with the initial
strategy allocation -- becomes increasingly important and
eventually leads to instability of the Eulerian Trail
quasi-attractor \cite{Jefferies1}.

Johnson {\em et al.} \cite{zheng2} subsequently introduced and
studied numerically what is known as the Binary-Agent-Resource
(B-A-R) model, in which the winning group is not necessarily
decided by the minority rule. The B-A-R model features a cutoff
parameter $L$, which is referred to as the global resource level
$L$ ($L < N$).  The values of $L$ and $N$ are not known to the
agents. At each timestep $t$, each agent decides upon two possible
options: whether to access the resource or not.  The winning
action is decided by whether the number of agents attempting to
access the resource, actually exceeds this resource level. The MG
therefore corresponds to the particular case of $L=N/2$ in the
B-A-R model. In Ref.\cite{zheng2}, it was found numerically that
the population may unwittingly self-segregate itself into groups
if $L$ deviates sufficiently from $N/2$. As an example, consider a
very high resource level of $L \approx N$  with each agent holding
two strategies: it was found numerically that approximately $3N/4$
agents are persistent winners while the rest are persistent
losers.

In the present work we analyze, both numerically and analytically,
the transition from modest to high resources in this generic B-A-R
model of a competing multi-agent population. Surprisingly, we find
that the system exhibits a set of abrupt transitions between
distinct yet well-defined states as the resource level $L$
increases from $N/2$ to $N$. In particular, both the highest
success rate $w_{max}$ and the mean success rate $\langle w
\rangle$ among the agents, show abrupt transitions as $L$ varies.
In addition, $w_{max}$ exhibits {\em fractional} values in the
plateau regions between each transition. We show that this
behavior can be understood in terms of the system's trajectory in
the history space as time evolves. In particular, as $L$
increases, the portion of the history space that the system visits
becomes increasingly restricted.  We derive analytic expressions
for the observed plateaux values and the values of $L$ at which
the transitions occur. Although the present analysis focuses on a
non-networked population, the same elements of (i) strategy
performance over time, and hence the dynamics of strategy scores,
plus (ii) the system's trajectory in history space, provide the
foundation for a quantitative understanding of a large class of
agent-based models, including networked populations
\cite{Lo2004,gourley}. We also note that for physicists interested
in random walks, the present B-A-R system provides a fascinating
laboratory for studying correlated, non-Markovian diffusion on a
non-trivial network (i.e. the history space, which corresponds to
a de Bruijn graph). This non-trivial diffusion is in turn strongly
coupled to the non-random temporal patterns arising in the
strategy-performance dynamics. Finally, we note in passing that
the occurrence of abrupt transitions between plateau states, and
stable fractions, are known to arise in a multi-electron quantum
system as the external magnetic field is increased monotonically
(i.e. Fractional Quantum Hall Effect \cite{FQHE}). However it is
very curious to see such `quantized' phenomena arise in a
classical multi-particle system as a function of a monotonically
increasing external control parameter.

The plan of the paper is as follows.  In Sec.~II, we define the
B-A-R model.  In Sec.~III, we present numerical results from
extensive simulations. In particular, we demonstrate the existence
of different phases or states at different values of the resource
level. In Sec.~IV, we discuss how the strategies' performance
evolves as the system evolves. In Sec.~V, we discuss the history
space of the B-A-R model and present results for the statistics in
the outcome bit-strings at high resource levels.  In Sec.~VI, we
explain the observed numerical features based on the idea that at
high resource levels, the system restricts itself to only visit a
restricted portion of the history space.  We derive an expression
for the highest success rate $w_{max}$ among the agents and
discuss the critical values of resource level at which transitions
occur.  We summarize the results in Sec.~VII, together with a
discussion of how the present approach can be generalized to a
wider class of agent-based models.

\section{Model}
\label{sec:PhaseTrans_Model}

We consider the Binary-Agent-Resource (B-A-R) model
\cite{johnson2,johnsonDPG,zheng2}. The B-A-R model is a binary
version of Arthur's El Farol bar attendance model
\cite{arthur1,johnson1}, in which a population of agents
repeatedly decide whether to go to a bar with limited seating
based on the information of the crowd size in recent weeks. In the
B-A-R model, there is a global resource level $L$ which is not
announced to the agents, where $N$ is the total number of agents.
At each timestep $t$, each agent decides upon two possible
options: whether to access resource $L$ (action `1') or not
(action `0'). The two global outcomes at each timestep, `resource
over-used' and `resource not over-used', are denoted by `$0$' and
`$1$'.  If the number of agents $n_{1}(t)$ choosing action $1$
exceeds $L$ (i.e. resource over-used and hence global outcome
`$0$') then the $N - n_{1}(t)$ abstaining agents win. By contrast
if $n_{1}(t) \leq L$ (i.e. resource not over-used and hence global
outcome `$1$') then the $n_{1}(t)$ agents win. In order to
investigate the behavior of the system as $L$ changes, it is
sufficient to study the range $N/2 \leq L \leq N$. The results for
the range $0 \leq L \leq N/2$ can be obtained from those in the
present work by suitably interchanging the role of `0' and `1'
\cite{zheng2}. In the special case of $L = N/2$, the B-A-R model
reduces to the Minority Game.

In the B-A-R model, each agent shares a common knowledge of the
past history of the most recent $m$ outcomes, i.e. the winning
option in the most recent $m$ timesteps. The full strategy space
thus consists of $2^{2^m}$ strategies, as in the MG. Initially,
each agent randomly picks $s$ strategies from the pool of
strategies, with repetitions allowed.  The agents use these
strategies throughout the game. At each timestep, each agent uses
his momentarily best performing strategy with the highest
virtual points. The virtual points for each strategy indicate the
cumulative performance of that strategy: at each timestep, one virtual
point (VP) is awarded (deducted) to (from) a strategy that would have
predicted the correct (incorrect) outcome after all decisions have been
made.  A random coin-toss is used to break ties between strategies. In
the B-A-R model, the population may or may not contain network
connections.  In the case of a networked population
\cite{gourley,Lo2004,LosAlamos,choe1} each agent has access to
additional information from his connected neighbors, such as his
neighbors' strategies and/or performance. In the present work, we
focus on the B-A-R model with a non-networked
population.

To evaluate the performance of an agent, one (real) point is
awarded to each winning agent at a given timestep. A maximum of
$L$ points per turn can therefore be awarded to the agents. An
agent has a success rate $w$, which is the mean number of points
awarded to the agent per turn over a long time window. The mean
success rate $\langle w \rangle$ among the agents is then defined
to be the mean number of points awarded per agent per turn, i.e.
an average of $w$ over the agents. We are interested in
investigating the details of how the success rate changes as the
resource level $L$ varies in the efficient phase, where the number
of strategies (repetitions counted) in play is larger than the
total number of distinct strategies in the strategy space.

\section{Numerical Results: Resource-driven states in B-A-R Model}
\label{sec:PhaseTrans_Phases}

The effects of varying $L$ were first reported by
Johnson {\em et al.} \cite{zheng2}.  These authors studied numerically
the dependence of the fluctuations in the number of agents taking a
particular option, on the memory size $m$ for different values of $L$.
For the MG (i.e. $L = N/2$) in the efficient phase (i.e. small values of
$m$) the number of agents making a particular choice varies from timestep
to timestep, with additional stochasticity introduced via
the random tie-breaking process. The corresponding period depends on the
memory length $m$. The underlying reason is that in the efficient
phase for $L=N/2$, no strategy is better overall than any other.
Hence there is a tendency for the system to restore itself after a
finite number of timesteps, thereby preventing a given strategy's VPs
from running away from the others. As a result, the outcome bit-string
shows the feature of anti-persistency or double periodicity
\cite{challet3,challet4,challet5,challet6,savit,Jefferies1,zheng1}.
Since a maximum of $L=N/2$ points can be awarded per turn, the
mean success rate $\langle w \rangle$ over a sufficiently large
number of timesteps is bound from above by $L/N = 1/2$.

In the B-A-R system with high resource level, the mean success
rate behaves differently.  Taking the extreme case of $L \simeq
N$, the winning action is obviously `1' (i.e. access resource) and in
principle every agent could win in every timestep.  The history
upon which the agents decide, is persistently $m$-bits of `1'.
However, due to the random initial strategy distribution, some
agents may not hold a strategy that predicts the winning option
for a history of $m$ `$1$'s. Therefore, there are still
losers and $\langle w \rangle$ is less than $L$. The number of
losers depends on $s$, the number of strategies that each agent
holds. For $s=2$ and assuming that the strategies are picked
randomly, a mean number of $N/4$ agents in a large population will
hold two strategies both predicting the wrong option.  The mean
success rate is thus given by $\langle w \rangle = 3/4$.

\begin{figure}[ht]
\includegraphics[width=.7\textwidth]{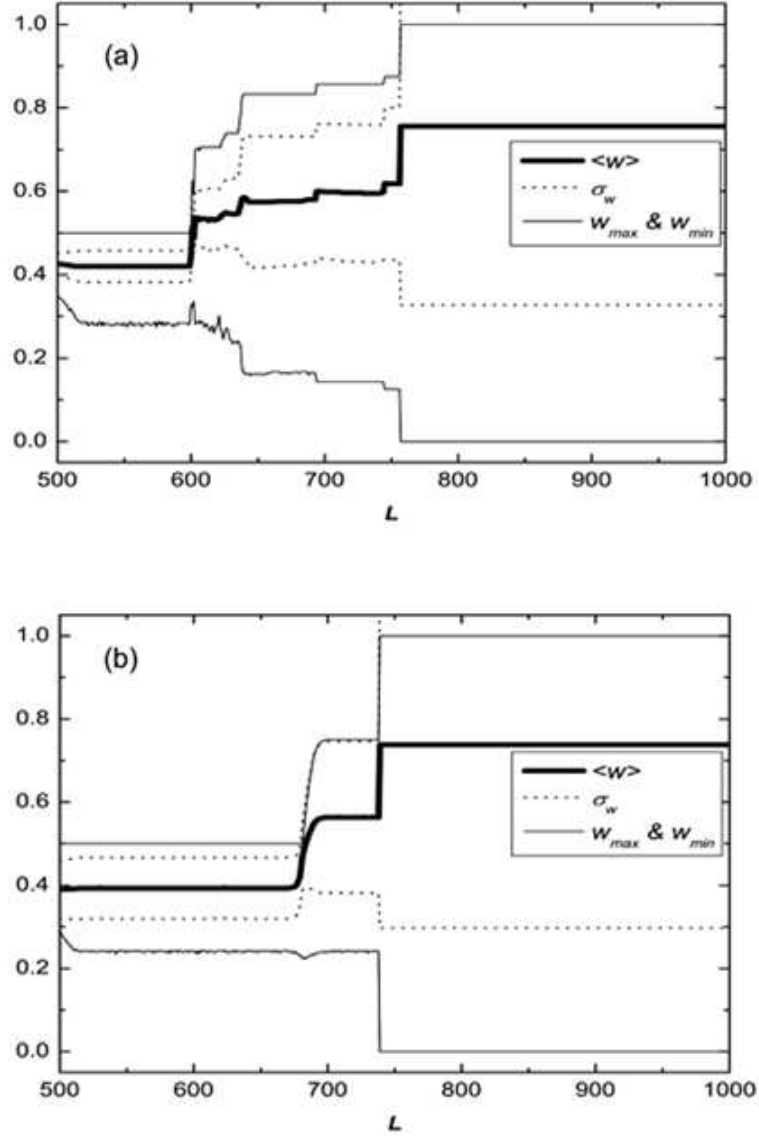}
\caption{The
mean success rate $\langle w \rangle$ (dark solid line) as a
function of the resource level $L$ for a system with $N=1001$
agents, $s=2$ strategies per agent, with memory length (a) $m=3$
and (b) $m=1$. For each value of $m$, data for different values of
$L$ are taken in a system with the same initial distribution of
strategies among the agents.  Also shown are the range
corresponding to one standard deviation about $\langle w \rangle$
in the success rates among the $N$ agents (dotted lines) and the
spread in the success rates given by the highest and the lowest
success rates (thin solid lines) among the agents.}
\label{fig:figure1}
\end{figure}

We have carried out extensive numerical simulations on the B-A-R
model to investigate the dependence of the success rate on $L$ for
$N/2 \leq L \leq N$.  Unless stated otherwise, we consider systems
with $N=1001$ agents and $s=2$. Figure~\ref{fig:figure1}(a) shows
the results of the mean success rate (dark solid line) as a
function of $L$ in a typical run for $m=3$, together with the
range corresponding to one standard deviation about $\langle w
\rangle$ in the success rates among the $N$ agents (dotted lines)
and the spread in the success rates given by the highest and the
lowest success rates (thin solid lines) in the population. By
taking a larger value of $N$ than most studies in the literature,
we can analyze the dependence on $L$ and $m$ in great detail.  In
particular, these quantities all exhibit abrupt transitions (i.e.
jumps) at particular values of $L$. Between the jumps, the
quantities remain essentially constant and hence form steps or
`plateaux'.  We refer to these different plateaux as states or
phases, since it turns out that the jump occurs when the system
makes a transition from one type of state characterizing the
outcome bit-string to another. For different runs, the results are
almost identical. At most, there are tiny shifts in the $L$ values
at which jumps arise due to (i) different initial strategy
distributions among the agents in different runs, and (ii)
different random initial history bit-strings used to start the
runs.

These different states are most clearly seen by monitoring the
{\em highest} success rate $w_{max}$ among the agents for given
values of $L$ and $m$.  The most striking feature in
Fig.\ref{fig:figure1}(a) is that the values of the plateaux in
$w_{max}$ are given by {\em simple fractions}, e.g. $7/8$, $6/7$,
$5/6$, $12/17$, $1/2$, etc.  This feature strongly suggests that
the system goes through different states with different ratios of
`$1$' and `$0$' in the outcome bit-string as $L$ varies, as will
be discussed in later sections. Figure~\ref{fig:figure1}(b) shows
that the features in the success rates for the simpler case of
$m=1$ are similar to those in Fig.\ref{fig:figure1}(a), except that
the plateaux in $w_{max}$ take on fewer values, i.e. $1$, $3/4$,
$1/2$ as $L$ decreases. These values are closely related to the
statistics in the outcome bit-string. For large $N$ and $m=1$, the
outcome bit-string shows a period of 4 bits. For values of $L$
with $w_{max} = 1/2$, it turns out that the fraction of the
outcome `$1$' in a period is exactly $1/2$. For the range of $L$
corresponding to $w_{max} = 3/4$, there are three $1$'s in a
period of $4$, and so on.  For $m=3$, we have also carried out
detailed analysis of the outcome bit-string. For later
discussions, we summarize in Table \ref{tab:otsm3s2} the values
of $w_{max}$, the range of $L$ corresponding to the observed value
of $w_{max}$, the ratio of number of occurrence of `1'-bits to
`0'-bits and the period in the outcome bit-string, as obtained
numerically from the data shown in Fig.\ref{fig:figure1}.
Hereafter, $w_{max}$ is used to label the state at a given $L$.

\section{Agents' Decisions and Strategy Performance}
\label{sec:PhaseTrans_Decision}

The agents decide based on the best performing strategy that they
hold at the moment of their decision.  A strategy's performance is
evaluated by its virtual points, which vary as the game proceeds.
It is most illustrative to consider the case of $m=1$ and $s=2$
(see Fig.\ref{fig:figure1}(b)) since one can readily follow the
dynamics for different values of $L$. For $m=1$, there are only
four different strategies in the whole strategy space.  These
strategies can be represented by ($00$), ($01$), ($10$), and
($11$), with the first (second) index in ($xy$) giving the action
for the history bit-string of `$0$' and `$1$', respectively. The
virtual points (VP) of the four strategies can then be represented
by a $2 \times 2$ matrix VP$_{xy}$, with the $xy$-element giving
the VP of the strategy ($xy$). Table~\ref{tab:strgevom1s2} shows
the time evolution of the VPs of the strategies in a few
timesteps, the number of agents $n_{1}$ taking the action `1', and
the outcome for $m=1$ in a population with {\em uniform} initial
distribution of all possible pairs of strategies to the agents. We
illustrate these ideas by considering the case of an initial
history of `0' for given $L$, but the same results are obtained
for an initial history of `1'. For $N/2 \leq L < 11N/16$, the
system follows the dynamics shown in the left column.  In this
range of $L$, the VPs cannot run away due to the decision making
process of the agents. The VPs restore their values in a few
timesteps, as in the MG.  For systems of large $N$, the outcome
series shows a 4-bit periodic pattern of $1100$, with $w_{max} =
1/2$ in agreement with the numerical results in
Fig.~\ref{fig:figure1}(b). This highest success rate is achieved by
the agents who hold two identical strategies. The result also
implies that an agent's success rate is determined by the Hamming
distance between the two strategies that the agent holds
\cite{Lopreprint}. One could therefore say that different
`species' of agent emerge in the population due to the dynamics of
the system, with each species characterized by its own Hamming
distance `gene'.

For $11N/16 \leq L < 3N/4$, two of the four strategies will have
runaway VPs, with one tending to increase without bound while
another tends to decrease without bound (see central column in
Table~\ref{tab:strgevom1s2}). Following the dynamics, the outcome
series shows a 4-bit periodic pattern of $1110$, in the limit of
large $N$. In this range, $w_{max} = 3/4$ as observed numerically
and this value is achieved by those agents holding the strategy
whose VPs increase without bound. For $L > 3N/4$, two strategies
have their VPs increasing (decreasing) without bound (see right
column in Table~\ref{tab:strgevom1s2}). The outcome series is
persistently `$1$', i.e. it becomes effectively period-1. In this
high resource level regime, $\langle w \rangle = 3/4$ since
three-quarters of the agents hold at least one strategy which
predicts the persistently winning option. In this way, it is
possible to follow the dynamics of the system and obtain the
number of possible states and the range of $L$ for each state.
Table~\ref{tab:wrm1s2} summarizes the theoretical results for
$m=1$ and $s=2$, by following the analysis on the dynamics as
shown in Table~\ref{tab:strgevom1s2}. The results for the
location of the transitions, the values of $\langle w \rangle$ and
$w_{max}$ are all in good agreement with numerical data (see
Fig.~\ref{fig:figure1}(b)). The important point is that for small
$m$, the system undergoes several changes of state with
successively higher values of $\langle w \rangle$ and $w_{max}$ as
$L$ increases. The value of $w_{max}$ is related to the ratio of
occurrences of the two possible outcomes in the outcome series,
which in turn is related to the strategies' performance and thus
the decision-making process for a given value of $L$.

We have carried out similar analysis for the case of $m=2$
and $m=3$. For $m=3$, the fraction of `$1$' in a period is found
to take on values in the set $\{\frac{8}{16}, \frac{12}{17},
\frac{17}{23}, \frac{5}{6}, \frac{6}{7}, \frac{7}{8}, 1\}$, which
coincides with the numerical values of $w_{max}$ obtained by
numerical simulations (see Fig.~\ref{fig:figure1}(a)) and shown in
Table~\ref{tab:otsm3s2}.  However, the analysis becomes
increasingly complicated for higher values of $m$ and/or $s$. The
reason is that there are $2^{2^{m}}$ strategies, and allowing $s$
strategies per agent divides the agents into $(2^{2^{m}})^{s}$
groups according to the sequence in which an agent picks his $s$
strategies.  This number increases rapidly with $m$ and $s$, and
the above microscopic analysis becomes hard to implement. It turns
out the states are closely related to the way in which the system
explores the possible histories.  In what follows, we analyze the B-A-R
model by an approach that focuses on the transitions between history
bit-strings and hence on the path in the history space
\cite{Jefferies1,Lo2004}.

\section{History space and bit-string statistics}
\label{sec:PhaseTrans_HistorySpace}

\subsection{History space}
Our approach couples together consideration of the probability of
the occurrence of various histories and the ranking in the
performance of the strategies \cite{Lo2004}. The history space
consists of all the possible history bit-strings for a given value
of $m$. For $m=3$, it includes $2^{3}$ bit-strings of $0$'s and
$1$'s. Figure~\ref{fig:figure2}(a) shows the history space for
$m=3$, together with the possible transitions from one history to
another.  Each history constitutes a node in the history space.
The transitions are marked by arrows together with the outcome
necessary for making the transitions. It will prove convenient to
group the possible history bit-strings for a given $m$ into
columns, in the way shown in Fig.~\ref{fig:figure2}(a). Each column
is labelled by a parameter $\zeta$, which is the number of `$0$'s
in the $3$-bit history (histories) concerned. One immediate
advantage of this labelling scheme is that the different states
characterized by $w_{max}$ turn out to involve paths in a {\em
restricted} portion of the full history space. For example, the
state with $w_{max}=1$ is restricted to the $\zeta = 0$ portion of
the history space, i.e. the $111$ history bit-string leads to an
outcome of `1' and hence persistent self-looping at the node $111$
in history space. The states with $w_{max}=7/8$, $6/7$, and $5/6$
correspond to different paths in the history space restricted to
the $\zeta=0$ and $\zeta=1$ groups of histories, as shown in
Fig.~\ref{fig:figure2}(b). The states with $w_{max}=17/23$ and
$12/17$ have paths extended to include $\zeta=2$ histories. The
state with $w_{max}=1/2$ has paths that cover the whole history
space ($\zeta = 0,1,2,3$).  In general, the deviation of $L$ from
$N/2$ acts like a driving force in the history space that drifts
the system towards an increasingly restrictive portion of the
history space bounded by a smaller value of $\zeta$. One can also
view this behavior as the system, in response to the global
resource level $L$,  effectively avoiding certain nodes in the
history space and hence avoiding particular patterns of historical
outcomes.

\subsection{Bit-string statistics of different states}
\label{sec:PhaseTrans_00states}

As the game proceeds, the system evolves from one history
bit-string to another.  This can be regarded as transitions
between different nodes (i.e. different histories) in the
history space. For $L=N/2$ in the efficient phase, it has been shown
\cite{savit} that the conditional probability of an outcome of,
say, `$1$' following a given history is the same for all
histories.  For $L \neq N/2$, the result still holds for states
characterized by $w_{max}=1/2$. Note that a history bit-string can
only make transitions to history bit-strings that differ by
the most recent outcome, e.g. 111 can only be make transitions to
either 110 or 111, and thus many transitions between two nodes in
the history space are forbidden. In addition, these allowed
transitions do not in general occur with equal probabilities. This leads
to specific outcome (and history) bit-string statistics for a state
characterized by $w_{max}$.

We have carried out detailed analysis of the statistics of the
outcomes following a given history bit-string for $m=3$, and for
each of the possible states over the whole range of $L$.
Table~\ref{tab:tmm3} gives the relative numbers of occurrences of
each outcome for every history bit-string.  For the state with
$w_{max} = 1/2 = 8/16$, for example, the outcomes `$0$' and `$1$'
occur with equal probability for every history bit-string, as in
the MG.  For the other states, the results reveal several striking
features. It turns out that $w_{max}$ is given by the {\em
relative frequency} of an outcome of `$1$' in the outcome
bit-string, which in turn is governed by the resource level $L$.
For example, a `0' to `1' ratio of $5:12$ in the outcome
bit-strings corresponds to the state with $w_{max}=12/17$. In
Table~\ref{tab:tmm3}, we have intentionally grouped the history
bit-strings into rows according to the label $\zeta$ in
Fig.~\ref{fig:figure2}.  We immediately notice that for every
possible state in the B-A-R model, the relative frequency of each
outcome is a property of the {\em group} of histories having the
same label $\zeta$ rather than the individual history bit-string,
i.e. all histories in a group have the same relative fraction of a
given outcome. This observation is important in understanding the
dynamics in the history space for different states in that it is
no longer necessary to consider each of the $2^{m}$ history
bit-strings in the history space. Instead, it is sufficient to
consider the four groups of histories (for $m=3$) as shown in
Fig.~\ref{fig:figure2}(a). Analysis of results for higher values of
$m$ show the same feature.

For the state characterized by $w_{max}=1$, the outcome bit-string
is persistently `$1$' and the path in the history space is
repeatedly 111$\rightarrow$1.  Therefore, the path is restricted
to the history labelled by $\zeta=0$ and simply corresponds to an
infinite number of loops around the history node 111. Since there
is no `$0$' in the outcome bit-string, we will also refer to this
state as $\zeta_{max}=0$ state. The system is effectively frozen
into one node in the history space. In this case, there are
effectively only two kinds of strategies, which differ by their
predictions for the history 111. The difference in predictions for
the other ($2^{m} -1$) history bit-strings become irrelevant.
Obviously, the ranking in the performance of the two effective
groups of strategies is such that the group of strategies that
suggest an action `1' for the history `111', outperforms the group
that suggests an action `0'. For a uniform initial distribution of
strategies, there are $N/2^{s}$ agents taking the action 0 and
$(1- 1/2^{s})N$ agents taking the action `1', since half of the
strategies predict 0 and half of them predict 1. To sustain a
winning outcome of `1', the criterion is that the resource level
$L$ should be higher than the number of agents taking the action
`1'. Therefore, we have for the state with $w_{max} =1$ that
\begin{equation}
 \langle w \rangle =  1-\frac{1}{2^s} \label{eq:wmean_00}
\end{equation}
and
\begin{equation}
 L  > \Big(1-\frac{1}{2^s}\Big)N. \label{eq:Lc_00}
\end{equation}
These results are in agreement with the results obtained by numerical
simulations. For $s=2$, $\langle w \rangle=3/4$ for $L > 3N/4$.
Note that Equations~\eqref{eq:wmean_00} and \eqref{eq:Lc_00} are
valid for {\em any} values of $m$.

Table~\ref{tab:tmm3} shows that the states with $w_{max} = 5/6$,
$6/7$, $7/8$ have very similar features in terms of the bit-string
statistics.  They differ only in the frequency of giving an
outcome of $1$ following the history of $111$.  Note that the
$\zeta = 2$ and $\zeta=3$ histories do not occur.  The results
imply that as the system evolves, the path in history space for
these states is restricted to the two groups of histories labelled
by $\zeta = 0$ and $\zeta=1$. The statistics show that the outcome
bit-strings for the states with $w_{max}=5/6$, $6/7$ and $7/8$
exhibit only one $0$-bit in a period of $6$, $7$ and $8$ bits,
respectively. We refer to these states collectively as
$\zeta_{max}=1$ states, since the portion of allowed history space
is bounded by the $\zeta=1$ histories.  Graphically, the path in
history space consists of a few self-loops at the node 111, i.e.
from 111 to 111, then passing through the $\zeta=1$ group of
histories once and back to 111, as shown in
Fig.~\ref{fig:figure2}(b). The states with $w_{max} = 1/2$, $12/17$,
$17/23$ involve the other groups of histories and exhibit
complicated looping among the histories. We refer to them
collectively as higher (i.e. $\zeta_{max} > 1$) states.

\section{The $\zeta_{max}=1$ states}
\label{sec:PhaseTrans_10states}

\subsection{Values of $w_{max}$}

We now proceed to derive an expression for the observed value of
$w_{max}$ for the $\zeta_{max}=1$ states.  Recall that each
strategy consists of a prediction or action for all of the $m$-bit
histories.  An important idea is that for states corresponding to
paths restricted to a certain portion of the history space, only
that part of a strategy corresponding to the histories in question
is being used in making decisions.  Strategies that only differ in
their predictions for the history bit-strings which {\em do not}
occur (i.e. the avoided histories) are now effectively identical.
In the context of the Crowd-Anticrowd theory
\cite{johnson1,johnson2,johnson3}, two previously uncorrelated
strategies could now be correlated when viewed within this
restricted history subspace.

\begin{figure}[ht]
\includegraphics[width=.7\textwidth]{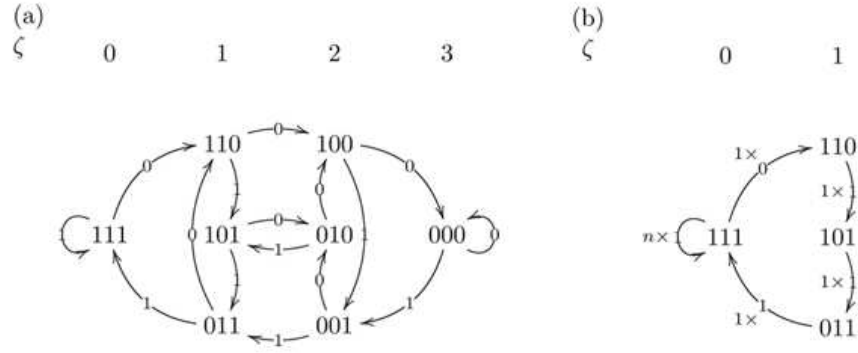}
\caption{(a) The
history space in B-A-R model with $m=3$. The nodes correspond to
the $2^{m}$ possible histories.  The transition between nodes are
indicated by the arrows, together with the outcome needed for the
transitions to occur.  The histories can be grouped into columns
labelled by a parameter $\zeta$ which gives the number of `0' bits
in the histories. (b) For the $\zeta_{max}=1$ states at $L
\lesssim (1- 1/2^{s})N$, the system follows a path restricted to
histories in the $\zeta=0$ and $\zeta=1$ columns, with $n$ loops
at the $\zeta=0$ history.} \label{fig:figure2}
\end{figure}

As $L$ decreases, the system is allowed to explore a larger
portion of the history space. The ranking in the strategies'
performance becomes more complicated.  For the $\zeta_{max}=1$
states, the paths in history space involve the ($m+1$) histories
labelled by $\zeta=$0 and 1 (see Fig.~\ref{fig:figure2}(b)). For
$m=3$, only four out of a total of $2^{m}=8$ entries in a strategy
corresponding to the histories `$111$', `$011$', `$110$', and
`$101$' now matter. For general $m$, there are $m$ histories with one
`0'-bit: hence a total of ($m+1$) entries in each strategy now
matter. For later discussions, it is useful to first classify all
strategies into two groups according to their prediction for the
history `11\dots1' ($\zeta=0$ history). These strategies can
further be classified according to their $m$ predictions for the
$\zeta=1$ histories. A strategy having $i$ bits predicting 1
and $m-i$ bits predicting 0 for the $m$ histories belonging to
$\zeta=1$, can be labelled as ($\mu$; $i$, $m-i$) where $\mu=0,1$
is the prediction for the $\zeta=0$ history. The first three
columns in Table~\ref{tab:strategies10} show this classification
of strategies for general values of $m$ describing the
$\zeta_{max}=1$ states.

For $L \lesssim 3N/4$, the outcomes are no longer persistently
`$1$', and the system explores both the $\zeta=0$ and $\zeta=1$
groups of histories. Assume that for $L$ just below $3N/4$, the
outcome must be `$1$' for the $\zeta=1$ histories, i.e. the system
only visits the $\zeta=1$ histories once in a cycle. Consider the
$m=3$ case, for example. As time evolves, the system exhibits
periodic visits to the histories. In each period, each history in
$\zeta =1$ occurs once and the history in $\zeta =0 $ occurs $n+1$
times. Among these $n+1$ occurrences of the $\zeta=0$ history, the
outcomes are `$1$' for $n$ timesteps and `$0$' for one timestep
(Fig.~\ref{fig:figure2}(b)), since the system must go from `$111$'
to `$110$' after $n$ loops in order to sustain the path. It turns
out that paths in the history space for the $w_{max} = 5/6$,
$6/7$, and $7/8$ states ($m=3$) correspond to that shown in
Fig.~\ref{fig:figure2}(b) with $n=2$, $3$, and $4$ loops at the
$\zeta=0$ history.

Since only $m+1$ different histories are involved, the performance
of a strategy depends only on the predictions for this subset of
histories.  Consider the path in Fig.~\ref{fig:figure2}(b). The
strategy labelled by $(\mu; i, m-i)$ predicts the correct outcome
$i+n$ times for $\mu=1$ and $i+1$ times for $\mu=0$, respectively,
in going through the path once. In Table~\ref{tab:strategies10},
we list the performance of the strategies labelled by
$(\mu;i,m-i)$ according to the number of successful predictions
(which reflects the VPs) in a closed path (see
Fig.~\ref{fig:figure2}(b)) consisting of $n$ loops at $\zeta=0$
history, i.e. a total of $(m+n+1)$ timesteps from $n=1$ to
$n=m+2$. It is important to note that there may be overlaps in
strategies' performance between the $\mu=1$ and $\mu=0$ groups of
strategies for small values of $n$, i.e. strategies with the label
$\mu=1$ and $\mu=0$ may win the same number of timesteps in a
cycle and hence belong to the same rank in performance of the
strategies. For example for $m=3$ and $n=2$, strategies labelled
by $(1;2,1)$ and $(0;3,0)$ belong to the same rank in performance.

The number of turns $n$ around the $\zeta=0$ history which is
consistent with the condition of $L < 3N/4$, is restricted to the
range $2 \leq n \leq m+1$.  This criteria is related to the number
$\tau$ of overlapping performances between the $(1;i,m-i)$ and
$(0;j,m-j)$ groups of strategies (see
Table~\ref{tab:strategies10}). Note that for $n > m+1$, we have $\tau
=0$, i.e. strategies with $\mu=1$ do not have overlapping VPs
with strategies with $\mu=0$. This implies that the VPs of the
strategies predicting `$1$' for the history in $\zeta =0$, are
always higher than those predicting `$0$'. This further implies
that agents will take action `$1$' for the $\zeta=0$ history if
one of their strategies belongs to the $\mu=1$ category.  For
$s=2$, there will then be $3N/4$ agents taking the action `$1$'.
Since $L<3N/4$, the outcome must be `$0$'. Therefore, the upper
bound for the number of self-loops is $m+1$, and thus $n \leq
m+1$. In other words, for paths with $n > m+1$ the system must
have $L > 3N/4$ and the path in history space will be that of an
infinite number of loops around $\zeta=0$ history, i.e. the
$\zeta_{max}=0$ state. For $n \leq 1$, strategies in $(1;i,m-i)$
will perform worse than or equally well to those in $(0;j,m-j)$
for $i>j$. As $L>N/2$, strategies that predict more $1$'s should
perform better. Therefore, $n \leq 1$ leads to inconsistency and
thus $n \geq 2$. Thus $2 \leq n \leq m+1$ for the $\zeta_{max}=1$
states, with the corresponding $\tau$ of overlapping groups of
strategies being in the range $1 \leq \tau \leq m$.  Since each
possible allowed value of $n$ or $\tau$ gives one $\zeta_{max}=1$
state, there are altogether $m$ possible $\zeta_{max}=1$ states
for a given $m$.

The values of $w_{max}$ for the $\zeta_{max}=1$ states can be
readily found. For a given value of $n$, the best performance
among the strategies is to have $m+n$ correct predictions in a
path consisting of ($m+n+1$) timesteps with $n$ loops at the
$\zeta=0$ history. Therefore, for the $\zeta_{max}=1$ states
\begin{equation}
 w_{max} = \frac{m+n}{m+n+1}. \label{eq:wmax_10}
\end{equation}
For $m=3$, we have $2 \leq n \leq 4$ and hence $n=2,3,4$. There
are three $\zeta_{max}=1$ states with $w_{max} = 5/6$, $6/7$, and
$7/8$, exactly as observed in the numerical simulations.

\subsection{Resource levels at transitions}

We now derive the critical values of the resource level at which
transitions occur from one value of $w_{max}$ to another for the
$\zeta_{max}=1$ states.  Note that the transition from the
$w_{max}=1$ state to the $w_{max} = 7/8$ state for $m=3$ and
$s=2$, occurs at $L=3N/4$ as predicted in Eq.(2).  The condition
for transitions from $w_{max}=7/8$ to $w_{max} = 6/7$ state is
that the value of $L$ can no longer support $n=4$ loops at the
$\zeta =0$ history before giving an outcome of `0' for the history
111. From Table~\ref{tab:strategies10}, we note that the
performance of the $\mu=1$ group of strategies becomes
increasingly better than the $\mu=0$ group as $n$ increases, and
the number of agents taking the action `1' increases towards
$3N/4$. Therefore, the highest number of agents who take the
action `1' {\em and} win will arise at the last turn among the $n$
loops where the history 111 is followed by an outcome `1'.
Similarly, the lowest number of winning agents will be at the turn
when the history 111 is followed by an outcome `0', i.e. breaking
away from the $n$ loops at 111.

The number of agents choosing the action `1' given the history 111, is
related to the number $\tau$ (and hence $n$) of overlapping performances
among the $(1;i,m-i)$ and $(0;j,m-j)$ strategies. From
Table~\ref{tab:strategies10}, the number of correct predictions
$v_{\mu}(i)$ of a strategy $(\mu;i,m-i)$ in going through a path
with $n$ loops at the $\zeta=0$ history, is given by
\begin{equation}
 v_{\mu}(i) = \left\{
  \begin{array}{ll}
   i+n & , \mu=1 \\
   i+1\phantom{M} & , \mu =0.
  \end{array} \right.
\end{equation}

For general values of $m$, there are only $(m+1)$ histories which matter
for the
$\zeta_{max}=1$ states.  Therefore there are $2^{m+1}$ {\em
effectively different} strategies, each of which represents a
group of $2^{2^{m}}/2^{m+1}$ strategies. The number of effectively
different strategies predicting $\mu$ for the $\zeta=0$ history
and having $i$ predictions of `1' for the $m$ $\zeta=1$ histories,
is given by
\begin{equation}
 \label{eq:cai}
 c_{\mu}(i) = \textrm{C}^{m}_{i}
\end{equation}
for both $\mu = 1$ and $0$, where $\textrm{C}^{m}_{i}$ is the
binomial coefficient.

The performance of the strategies can be ranked by a label $r$,
with $r=1,\dots,r_{max}$ and $r=1$ representing the best
performing group of strategies.  A general situation in which
there are $\tau$ overlapping performances between the $(1;i,m-i)$
strategies and $(0;j,m-j)$ strategies, is shown in
Table~\ref{tab:VPrank10}. For the $\zeta_{max}=1$ states, the
best performing strategies belong to the $\mu=1$ group and the
worst performing ones belong to the $\mu=0$ group, with $\tau$
overlapping rankings in between where the allowed range of $\tau$
is $1 \leq \tau \leq m$. The ranking $r$ of the strategies
$(\mu;i,m-i)$ is related to $i$ by the simple relation
\begin{equation}
 \label{eq:ir_relation}
 i = \left\{
  \begin{array}{ll}
   m+1-r & , \mu=1 \\
   2m+2-\tau-r\phantom{M} & , \mu=0.
  \end{array} \right.
\end{equation}
For a given value of $\tau$, there are a total of $2m+2-\tau$ ranks.
Therefore, the label $r$ is restricted to the range $1 \leq r \leq
r_{max}$ with $r_{max}$ given by
\begin{equation}
 \label{eq:rmax}
 r_{max} = 2m+2-\tau.
\end{equation}
As the number of loops $n$ increases, $\tau$ decreases and the
strategies spread more widely in terms of performance. It follows
from Eqs.~\eqref{eq:cai} and \eqref{eq:ir_relation} that the
number of effectively different strategies $c(r)$ in rank-$r$ is
given by
\begin{eqnarray}
 \label{eq:cr}
 c(r)
 & = & \left\{
  \begin{array}{ll}
   \textrm{C}^{m}_{r-1} & , \; r\in(1,m+1-\tau) \\
   \textrm{C}^{m}_{r-1} + \textrm{C}^{m}_{r_{max}-r} \phantom{M} & , \; r\in(m+2-\tau,m+1) \\
   \textrm{C}^{m}_{r_{max}-r} & ,\; r\in(m+2,r_{max}).
  \end{array} \right.
\end{eqnarray}
The fraction of rank-$r$ strategies among all the strategies is
then given by
\begin{equation}
\label{eq:tildec} \widetilde{c}(r) = \frac{c(r)}{2^{m+1}}.
\end{equation}

On the critical turn that determines the minimum value of $L$ for
sustaining a certain $\zeta_{max}=1$ state, only the $\mu=1$
strategies win. Using Table~\ref{tab:VPrank10} together with
Eqs.~\eqref{eq:cai} and \eqref{eq:ir_relation}, the number of
winning strategies $c_{1}(r)$ is
\begin{eqnarray}
 \label{eq:cvr}
 c_{1}(r)
 & = & \left\{
  \begin{array}{ll}
   \textrm{C}^{m}_{r-1}\phantom{M} & , r\in(1,m+1) \\
   0 & , r\in(m+2,r_{max}),
  \end{array} \right.
\end{eqnarray}
The corresponding fraction $f_{1}(r)$ of winning strategies among
all strategies of rank-r is given by
\begin{eqnarray}
 \label{eq:fvr}
 f_1(r)
 & = & \frac{c_1(r)}{c(r)} =  \left\{
  \begin{array}{ll}
   1 & , r\in(1,m+1-\tau) \\
   \frac{1}{1+\textrm{C}^{m}_{r_{max}-r}/
    \textrm{C}^{m}_{r-1}}\phantom{M}
    & , r\in(m+2-\tau,m+1) \\
   0 & , r\in(m+2,r_{max}).
  \end{array} \right.
\end{eqnarray}

Each agent uses the strategy in his possession which has the best
performance record, i.e. the one having the ranking with smaller
$r$, in order to make a decision.  Assuming a uniform initial
distribution of any combination of $s$ strategies (with
repetitions allowed) among the agents, the fraction of agents
\emph{holding} a rank-$r$ strategy as their best performing
strategy $\widetilde{n}_H(r)$ is
\begin{eqnarray}
 \label{eq:nhsr}
 \widetilde{n}_H(r) & = & \Bigg[\sum_{r'=r}^{r_{max}}\widetilde{c}(r')\Bigg]^s
  -\Bigg[\sum_{r'=r+1}^{r_{max}}\widetilde{c}(r')\Bigg]^s.
\end{eqnarray}
with $\widetilde{c}(r)$ given by Eq.\eqref{eq:tildec}.

Each $\zeta_{max}=1$ state corresponds to a specific value of
allowed $n$ and hence an allowed $\tau$. For a given $n$ or $\tau$,
the resource level $L$ needed to accommodate all the agents that
take the action `1' for the $\zeta=0$ history, gives the criterion
for the state:
\begin{eqnarray}
 \label{eq:Lc_10_s2}
 L & > & N\sum_{r=1}^{r_{max}}f_1(r)\widetilde{n}_H(r).
\end{eqnarray}
Note that $r_{max}$, $f_1(r)$, and $\widetilde{n}_H(r)$ are all
$\tau$-dependent (see Eqs.\eqref{eq:rmax}, \eqref{eq:fvr}, and
\eqref{eq:Lc_10_s2}).  Equation \eqref{eq:Lc_10_s2} gives the
lower bounds of $L$ for each of the $\zeta_{max}=1$ states. Note
that the lower bound for a state with a given $\tau$ is also the
upper bound for the state with $\tau+1$. For $N=1001$, $s=2$, and
$\tau=1$, $2$, and $3$, Eq.\eqref{eq:Lc_10_s2} gives the lower
bounds of $L=745$, $695$, and $640$ for the states characterized
by $w_{max} = 7/8$, $6/7$, and $5/6$, respectively. These values
are in excellent agreement with those obtained by numerical
simulations (see Fig.~\ref{fig:figure1}(a) and
Table~\ref{tab:otsm3s2}). We note that our approach of focusing
on strategy-performance ranking patterns and the fraction of
strategies in each rank, represents a generalization of a similar
approach \cite{Lo2004} that has already been successfully applied
to the MG to cases in which some of the strategies have runaway
VPs.

\section{Discussion}

We have studied numerically and analytically the effects of a
varying resource level $L$ on the success rate of the agents in a
competing population within the B-A-R model.  We found that
the system passes through different states, characterized either by
the mean success rate $\langle w \rangle$ or by the highest
success rate in the population $w_{max}$, as $L$ decreases from the
high resource level limit.  The number of states depends on
details of the system such as the memory size $m$ and the number
of strategies per agent $s$. Transitions between these states
occur at specific values of the resource level.  For small values
of $m$, it is possible to explain these states by following the
evolution of the performance of the strategies and the decision
making dynamics of the system.  More generally, we found that
different states correspond to different paths covering a
subspace within the whole history space. In the high resource level
regime, namely $L > (1-1/2^{s})N$, $w_{max}=1$. The corresponding
path in the history space is one that loops around the history
`111...' indefinitely.  Just below the high resource level regime
is a range of $L$ that gives $m$ states corresponding to the
fractions $w_{max} = (m+n)/(m+n+1)$. This result is in excellent
agreement with that obtained by numerical simulations.  For these
$\zeta_{max}=1$ states, i.e. the outcome series consists of one bit of
`0` in a cycle of $m+n+1$ bits, the path in history space is restricted to
those $m$-bit histories with at most one-bit of `0' and with $n$ loops
around the `111...' history. This identification of an active
portion within the history space implies that only part of each
strategy is being used.  By considering the performance of the
strategies within this restricted portion of the history space, the
number of loops $n$ consistent with the $\zeta_{max}=1$ states was
found to be $2 \leq n \leq m+1$.  The number of agents using a
strategy that predicts the action `1' given the history `111...'
increases as the number of loops $n$ increases.  Thus a criterion
on the resource level for sustaining a state of given $n$ can be
derived.  The results are again in excellent agreement with
numerical results. After passing through the $\zeta_{max}=1$
states, the system goes into states with more than one `0'-bit per
cycle in the outcome series as $L$ is further reduced. These
$\zeta_{max}>1$ states correspond to paths that explore an
increasingly larger portion of the history space. While our
analysis can also be applied to these $\zeta_{max}>1$ states, the
dynamics and the results are too complicated to be included here.

A resource level $L$ that deviates from $N/2$ acts like a {\em
driving force} in the history space. In response to this driving
force, the system effectively adjusts its dynamics to occupy an
increasingly restricted portion of the history space as $L$
increases.  For $L=N/2$, the system explores the whole history
space by passing through trails that are almost Eulerian
\cite{Jefferies1}.  The random initial distribution of strategies
and the random initial history bit-string that started the system,
provide the seed for the diffusive behavior which develops in the
history space as the system evolves.  In fact, results of
numerical simulations for $L \gtrsim N/2$ show that slightly
increasing $L$ beyond $N/2$ has the effect of suppressing this
random wandering through the history space, and instead locks the
system into the Eulerian Trail. However opposing mechanisms can
arise to counteract this driving force, thereby enhancing the
diffusive behavior. For example, this can be achieved by allowing
the agents the chance of using a strategy besides the
best-performing one \cite{thermal1,thermal2} or by allowing some
agents to opt out of the system occasionally \cite{larry1}.
Alternatively, the system can be biased through the initial
strategy scores, or by introducing a specially prepared non-random
initial allocation of strategies \cite{Jefferies1}. It is this
competition between the diffusive and driven behavior that gives
the non-trivial global behavior in the B-A-R model and its
variations. For this reason, the present B-A-R system provides a
fascinating laboratory for studying  correlated, non-Markovian
diffusion on a non-trivial network (i.e. history space). This
non-trivial diffusion is in turn strongly coupled to the
non-random temporal patterns arising in the VP dynamics. We note
that the present results could also be used to generalize the
Crowd-Anticrowd theory in order to incorporate the effect of
restricted history-space dynamics: this would then allow
identification of an appropriate set of correlated, uncorrelated
and anti-correlated strategies in order to implement the
Crowd-Anticrowd theoretical expressions.

As a side-product, our analysis serves to illustrate the
sensitivity within multi-agent models of competing populations, to
tunable parameters. By tuning an external parameter, which we take
as the resource level in the present work, the system is driven
through different paths in the history space which can be regarded
as a `phase space' of the system. The feedback mechanism, which is
built-in through the decision making process and the evaluation of
the performance of the strategies, makes the system highly
sensitive to the resource level in terms of which states the
system decides to settle in or around. These features are quite
generally found in a wide range of complex systems.  The ideas in
the analysis carried out in the present work, while specific to
the B-A-R model used, are also applicable to other models of
complex systems.

In closing, we remark that besides obtaining analytically the
highest success rate $w_{max}$ and the criteria on the resource
level $L$, our treatment can also be extended to obtain the mean
success rate $\langle w \rangle$ for the $\zeta_{max}=1$ states.
The analysis is more complicated than that for obtaining
$w_{max}$. The procedure is to follow the evolution of the
performance of the groups of strategies in {\em each timestep}
through a path in the history space.  The number of agents taking
a particular action, and hence the number of winning agents, can
be found from a strategies' performance table like the one shown
in Table~\ref{tab:strategies10}.  We have carried out the
analysis for $\langle w \rangle$ for $m=1$, $2$, and $3$, and
results are found to be in excellent agreement with numerical
results.  Our analysis can also be readily extended to consider
connected populations in which agents have established links to
connected neighbors for collecting additional information
\cite{gourley,choe1,Lo2004}. Quite generally, the effect of the
links is to modify the number of agents {\em using} a strategy in
a particular rank.  For connected populations, an agent may use a
strategy that he does not hold but has access to through his
links.  For an agent who uses the best performing strategy among
his own $s$ strategies and those of his connected neighbors, the
success rate behaves in a similar fashion as a function of $L$ as
that reported here, only that the critical values of resource
level at which transitions occur are shifted \cite{sonic}. These
results can be understood by incorporating the effects of the
linkages into Eq.\eqref{eq:nhsr}. Results for the B-A-R model in a
connected population will be reported elsewhere \cite{sonic}.

\acknowledgments This work was supported in part by the Research
Grants Council of the Hong Kong SAR Government through Grant No.
CUHK4121/01P.  We acknowledge useful discussions with Ho-Yin Lee,
Keven K. P. Chan, Charley S. Choe and Sean Gourley.


\newpage
\begin{table}[ht]
\begin{tabular}{|c||c|c|c|}
\hline
 $w_{max}$ & `1'-bits:`0'-bits in a period & Range of $L$ & \phantom{M}Period\phantom{M} \\
\hline \hline
 $1/2$ & 8:8 & $\sim$ 510-600 & length 16 \\
\hline
 $12/17$ & 12:5 & $\sim$ 600-620 & length 17 \\
\hline
 $17/23$ & 17:6 & $\sim$ 620-640 & length 23 \\
\hline
 $5/6$ & 5:1 & $\sim$ 640-695 & 111110 \\
\hline
 $6/7$ & 6:1 & $\sim$ 695-745 & 1111110 \\
\hline
  $7/8$ & 7:1 & $\sim$ 745-755 & 11111110 \\
\hline
 $1$ & 1:0 & $\sim$ 755-1000 & 1 \\
\hline
\end{tabular}
\caption{Table showing the states characterized by $w_{max}$ for
B-A-R model with $N=1001$ agents, $m=3$ and $s=2$, together with
the ratio of number of occurrences of `1'-bits to `0'-bits in the
outcome bit-string and the range of resource level in which the
state occurs.  The results are obtained from the numerical data as
shown in Fig.~\ref{fig:figure1}(a). The last column shows the
period observed in the outcome bit-string.} \label{tab:otsm3s2}
\end{table}

\clearpage
\begin{table}[ht]
\renewcommand{\arraystretch}{0.7}
\begin{tabular}{|c|c|c|c|}
 \multicolumn{1}{c}{} & \multicolumn{3}{c}{Initial history = 0} \\
\hline
 Timestep & \multicolumn{3}{c|}{$\Rightarrow$ outcome} \\
  & \multicolumn{3}{c|}{(VP)$_{(xy)}$   $n_{1}(t)$} \\
\hline
 & \multicolumn{3}{c|}{$\Rightarrow0$ (initial history)} \\
  1 & \multicolumn{3}{c|}{ $\binom{0\phantom{o}0}{0\phantom{o}0}$ $\frac{1}{2}N$} \\
 & \multicolumn{3}{c|}{$\Rightarrow1$} \\
  2 & \multicolumn{3}{c|}{ $\binom{-1\phantom{o}-1}{+1\phantom{o}+1}$ $\frac{1}{2}N$} \\
 & \multicolumn{3}{c|}{$\Rightarrow1$} \\
  3 & \multicolumn{3}{c|}{ $\binom{-2\phantom{o}0}{0\phantom{o}+2}$ $\frac{1}{2}N$} \\
 & $\Rightarrow0$ & \multicolumn{2}{c|}{$\Rightarrow1$} \\
  4 & $\binom{-1\phantom{o}-1}{+1\phantom{o}+1}$ $\frac{3}{4}N$
   & \multicolumn{2}{c|}{$\binom{-3\phantom{o}+1}{-1\phantom{o}+3}$ $\frac{3}{4}N$} \\
 & $\Rightarrow0$ & $\Rightarrow0$ & $\Rightarrow1$ \\
  5 & $\binom{0\phantom{o}0}{0\phantom{o}0}$ $\frac{1}{2}N$
   & $\binom{-2\phantom{o}0}{0\phantom{o}+2}$ $\frac{11}{16}N$
   & $\binom{-4\phantom{o}+2}{-2\phantom{o}+4}$ $\frac{3}{4}N$ \\
 & & $\Rightarrow1$ & \\
  6 & $\vdots$
   & $\binom{-3\phantom{o}-1}{+1\phantom{o}+3}$ $\frac{5}{8}N$
   & $\vdots$ \\
 & & $\Rightarrow1$ & \\
  7 &
   & $\binom{-4\phantom{o}0}{0\phantom{o}+4}$ $\frac{11}{16}N$
   & \\
 & & & \\
  8 &
   & $\vdots$
   & \\

 & & & \\
 \hline
 $L_{min}$ & $\frac{1}{2}N$
   & $\frac{11}{16}N$
   & $\frac{3}{4}N$ \\

 \hline \hline
  & \multicolumn{1}{p{2.1cm}|}
   {\scriptsize \setlength{\baselineskip}{0.5\baselineskip} VP
   ranking pattern eventually repeats in four timesteps, with a 4-bit period of 1100}
  & \multicolumn{1}{p{2.1cm}|}
   {\scriptsize \setlength{\baselineskip}{0.5\baselineskip} VP
   ranking pattern eventually repeats in four timesteps, with a 4-bit period of 1110}
  & \multicolumn{1}{p{2.1cm}|}
   {\scriptsize \setlength{\baselineskip}{0.5\baselineskip} VP
   ranking pattern eventually repeats in every timestep, with the outcome being
   persistently `1'} \\
 \hline
\end{tabular}
\caption{Time evolution of a B-A-R system for $m=1$ and $s=2$. The
virtual points (VP)$_{(xy)}$ of the strategies $(xy)$ are given for a few
timesteps, together with the number of agents $n_{1}(t)$ taking the
action `1' and the outcome of each timestep in the format given in the
first row of the table. Initially, the VPs of all strategies are set to
zero and a uniform initial distribution of strategies in a large $N$
population is assumed. The system settles into different states
depending on the resource level
$L$.} \label{tab:strgevom1s2}
\end{table}

\clearpage
\begin{table}[ht]
\begin{tabular}{|c||c|c|c|c|}
\hline
 $w_{max}$ & `1'-bits:`0'-bits in a period & Range of $L$ & \phantom{M}$\langle w \rangle$\phantom{M} \\
\hline \hline
 $\frac{2}{4}$ & 2:2 & 500-688 & $\frac{25}{64}$ \\
\hline
 $\frac{3}{4}$ & 3:1 & 688-751 & $\frac{9}{16}$ \\
\hline
 $1$ & 1:0 & 751-1000 & $\frac{3}{4}$ \\
\hline
\end{tabular}
\caption{Values of the mean success rates $\langle w \rangle$ for
states corresponding to different resource level $L$ in a B-A-R
model of $N=1001$ agents, $m=1$ and $s=2$, obtained analytically
by following the dynamics of the system as shown in
Table~\ref{tab:strgevom1s2}.  Results are in excellent agreement
with the simulation data given in Figure~\ref{fig:figure1}(b).}
\label{tab:wrm1s2}
\end{table}

\clearpage
\begin{table}[htb]
\begin{tabular}{|c|c|cc|}
\hline
 \multicolumn{2}{|c|}{$w_{max}=\frac{8}{16}$} & $\rightarrow$0 & $\rightarrow$1 \\
\hline \hline
 $\zeta=3$ & 000 & 1 & 1 \\
\hline
           & 001 & 1 & 1 \\
 $\zeta=2$ & 010 & 1 & 1 \\
           & 100 & 1 & 1 \\
\hline
           & 011 & 1 & 1 \\
 $\zeta=1$ & 101 & 1 & 1 \\
           & 110 & 1 & 1 \\
\hline
 $\zeta=0$ & 111 & 1 & 1 \\
\cline{1-4} \cline{3-4}
 \multicolumn{2}{c|}{} & 8 & 8 \\
\cline{3-4}
\end{tabular}
\begin{tabular}{|c|cc|}
\hline
 $\frac{12}{17}$ & $\rightarrow$0 & $\rightarrow$1 \\
\hline \hline
 000 & 0 & 0 \\
\hline
 001 & 0 & 1 \\
 010 & 0 & 1 \\
 100 & 0 & 1 \\
\hline
 011 & 1 & 2 \\
 101 & 1 & 2 \\
 110 & 1 & 2 \\
\hline
 111 & 2 & 3 \\
\cline{1-3} \cline{2-3}
 \multicolumn{1}{c|}{} & 5 & 12 \\
\cline{2-3}
\end{tabular}
\begin{tabular}{|c|cc|}
\hline
 $\frac{17}{23}$ & $\rightarrow$0 & $\rightarrow$1 \\
\hline \hline
 000 & 0 & 0 \\
\hline
 001 & 0 & 1 \\
 010 & 0 & 1 \\
 100 & 0 & 1 \\
\hline
 011 & 1 & 3 \\
 101 & 1 & 3 \\
 110 & 1 & 3 \\
\hline
 111 & 3 & 5 \\
\cline{1-3} \cline{2-3}
 \multicolumn{1}{c|}{} & 6 & 17 \\
\cline{2-3}
\end{tabular}

\vskip0.5cm

\begin{tabular}{|c|cc|}
\hline
 $\frac{5}{6}$ & $\rightarrow$0 & $\rightarrow$1 \\
\hline \hline
 000 & 0 & 0 \\
\hline
 001 & 0 & 0 \\
 010 & 0 & 0 \\
 100 & 0 & 0 \\
\hline
 011 & 0 & 1 \\
 101 & 0 & 1 \\
 110 & 0 & 1 \\
\hline
 111 & 1 & 2 \\
\cline{1-3} \cline{2-3}
 \multicolumn{1}{c|}{} & 1 & 5 \\
\cline{2-3}
\end{tabular}
\begin{tabular}{|c|cc|}
\hline
 $\frac{6}{7}$ & $\rightarrow$0 & $\rightarrow$1 \\
\hline \hline
 000 & 0 & 0 \\
\hline
 001 & 0 & 0 \\
 010 & 0 & 0 \\
 100 & 0 & 0 \\
\hline
 011 & 0 & 1 \\
 101 & 0 & 1 \\
 110 & 0 & 1 \\
\hline
 111 & 1 & 3 \\
\cline{1-3} \cline{2-3}
 \multicolumn{1}{c|}{} & 1 & 6 \\
\cline{2-3}
\end{tabular}
\begin{tabular}{|c|cc|}
\hline
 $\frac{7}{8}$ & $\rightarrow$0 & $\rightarrow$1 \\
\hline \hline
 000 & 0 & 0 \\
\hline
 001 & 0 & 0 \\
 010 & 0 & 0 \\
 100 & 0 & 0 \\
\hline
 011 & 0 & 1 \\
 101 & 0 & 1 \\
 110 & 0 & 1 \\
\hline
 111 & 1 & 4 \\
\cline{1-3} \cline{2-3}
 \multicolumn{1}{c|}{} & 1 & 7 \\
\cline{2-3}
\end{tabular}
\begin{tabular}{|c|cc|}
\hline
 $1$ & $\rightarrow$0 & $\rightarrow$1 \\
\hline \hline
 000 & 0 & 0 \\
\hline
 001 & 0 & 0 \\
 010 & 0 & 0 \\
 100 & 0 & 0 \\
\hline
 011 & 0 & 0 \\
 101 & 0 & 0 \\
 110 & 0 & 0 \\
\hline
 111 & 0 & 1 \\
\cline{1-3} \cline{2-3}
 \multicolumn{1}{c|}{} & 0 & 1 \\
\cline{2-3}
\end{tabular}

\caption{Outcome statistics of a B-A-R model with $N=1001$ agents,
$m=3$ and $s=2$ for states corresponding to different resource
level $L$.  The table shows the relative number of occurrence of
each outcome following every possible history bit-string for the
states characterized by $w_{max}$ = $8/16$, $12/17$, $17/23$,
$5/6$, $6/7$, $7/8$, and $1$.  The parameter $\zeta$ labels groups
of histories as defined in Fig.~\ref{fig:figure2}.}
\label{tab:tmm3}
\end{table}

\clearpage
\begin{table}[ht]
\begin{tabular}{||c|cc||p{1.5cm}p{1.5cm}p{1.0cm}p{1.8cm}p{1.0cm}p{1.5cm}p{1.5cm}||}
\hline

 \multicolumn{1}{||p{1.2cm}}{\setlength{\baselineskip}{0.5\baselineskip}
  $\zeta$=0 history} &
   \multicolumn{2}{|p{2.5cm}||}{\setlength{\baselineskip}{0.5\baselineskip}
    $\zeta$=1 histories} &
     \multicolumn{7}{|p{6.5cm}||}{\setlength{\baselineskip}{0.5\baselineskip}
      number of correct predictions in a closed path with $n$ loops at $\zeta$=0 history} \\
$\mu$ & $\rightarrow$1 & $\rightarrow$0 &
  \multicolumn{1}{c}{1} & \multicolumn{1}{c}{2} & $\cdots$ & \multicolumn{1}{c}{$n$}
   & $\cdots$ & \multicolumn{1}{c}{$m+1$} & \multicolumn{1}{c||}{$m+2$} \\
\hline \hline
 1 & $m$ & $0$ & $m+1$ & $m+2$ & $\cdots$ & $m+n$ & $\cdots$ & $2m+1$ & $2m+2$ \\
 1 & $m-1$ & $1$ & $m$ & $m+1$ & $\cdots$ & $m+n-1$ & $\cdots$ & $2m$ & $2m+1$ \\
 $\vdots$ & $\vdots$ & $\vdots$ & $\vdots$ & $\vdots$ &  & $\vdots$ &  & $\vdots$ & $\vdots$ \\
 1 & $i$ & $m-i$ & $i+1$ & $i+2$ & $\cdots$ & $n+i$ & $\cdots$ & $m+i+1$ & $m+i+2$ \\
 $\vdots$ & $\vdots$ & $\vdots$ & $\vdots$ & $\vdots$ &  & $\vdots$ &  & $\vdots$ & $\vdots$ \\
 1 & $0$ & $m$ & $1$ & $2$ & $\cdots$ & $n$ & $\cdots$ & $m+1$ & $m+2$ \\
\hline
 0 & $m$ & $0$ & $m+1$ & $m+1$ & $\cdots$ & $m+1$ & $\cdots$ & $m+1$ & $m+1$ \\
 $\vdots$ & $\vdots$ & $\vdots$ & $\vdots$ & $\vdots$ &  & $\vdots$ &  & $\vdots$ & $\vdots$ \\
 0 & $i$ & $m-i$ & $i+1$ & $i+1$ & $\cdots$ & $i+1$ & $\cdots$ & $i+1$ & $i+1$ \\
 $\vdots$ & $\vdots$ & $\vdots$ & $\vdots$ & $\vdots$ &  & $\vdots$ &  & $\vdots$ & $\vdots$ \\
 0 & $0$ & $m$ & $1$ & $1$ & $\cdots$ & $1$ & $\cdots$ & $1$ & $1$ \\
\hline
\end{tabular}
\caption{All strategies can be labelled by $(\mu;i,m-i)$ as shown
in the first three columns in the table. For paths with $n$ loops
as shown in Fig.~\ref{fig:figure2}(b), the number of correct
predictions by each group of strategies is given for different
values of $n$ from $n=1$ to $n=m+2$. For the $\zeta_{max}=1$
states, $2 \leq n \leq m+1$.} \label{tab:strategies10}
\end{table}

\clearpage
\begin{table}[ht]
\begin{tabular}{|c|cc|c|}
\cline{1-3}
  & \multicolumn{2}{c|}{$v_{\mu}(r)$} \\
 $r$ & $\mu=1$ & $\mu=0$ \\
\hline
 1 & $r_{max}$ & & \\
 2 & $r_{max}-1$ & & $m+1-\tau$ \\
 \vdots & \vdots & & ($\mu=1$ strategies only) \\
\hline
 \vdots & \vdots & \vdots & \\
 r & $r_{max}-r+1$ & $r_{max}-r+1$ & $\tau$ \\
 \vdots & \vdots & \vdots & (overlapping region) \\
\hline
 \vdots & & \vdots & \\
 \vdots & & 2 & $m+1-\tau$ \\
 $r_{max}$ & & 1 & ($\mu=0$ strategies only) \\
\hline
\end{tabular}
\caption{Table showing the strategy performance ranking pattern,
as reflected by the number of correct predictions $v_{\mu}(r)$, in
the $n$-th loop in a path given in Fig.~\ref{fig:figure2}(b) for
$\zeta_{max}=1$ states corresponding to turn with the largest
winning crowd.  Note that some $\mu=1$ groups of stategies may
have overlapping cumulative performance with $\mu=0$ groups of
strategies. The total number of ranks $r_{max}$ and the number of
$\mu=0$ and $\mu=1$ groups of strategies having identical rankings
depend on $n$ and $m$.} \label{tab:VPrank10}
\end{table}

\newpage

\newpage

\newpage

\end{document}